\documentclass[prl,twocolumn,amsmath,amsfonts,amssymb,showpacs]{revtex4} 
\usepackage{epsfig,psfrag,graphicx,bm,color}

\begin{document}
\title{Significant Gamma Lines from Inert Higgs Dark Matter}

\author{Michael Gustafsson}
\email{michael@physto.se}
\author{Erik Lundstr\"om}
\email{erik@physto.se}
\author{Lars Bergstr\"om}
\email{lbe@physto.se}
\author{Joakim Edsj\"o}
\email{edsjo@physto.se}

\affiliation{Department of Physics, Stockholm University, AlbaNova
 University Center, SE - 106 91 Stockholm, Sweden}

\date{October 4, 2007}


\begin{abstract}
One way to unambiguously confirm the existence of particle dark
matter and determine its mass would be to detect its annihilation
into monochromatic gamma-rays in upcoming telescopes. One of the
most minimal models for dark matter is the inert doublet model,
obtained by adding another Higgs doublet with no direct coupling to
fermions.
For a mass between 40 and 80 GeV, the lightest of the new inert
Higgs particles can give the correct cosmic abundance of cold dark
matter in agreement with current observations. We show that for this
scalar dark matter candidate, the annihilation signal of
monochromatic $\gamma\gamma$ and $Z\gamma$ final states would be
exceptionally strong. The energy range and rates for these gamma-ray
line signals make them ideal to search for with the soon upcoming
GLAST satellite. 
\end{abstract}

\maketitle

\newcommand{\nn}{\nonumber}
\newcommand{\ga}{\gamma}
\hyphenation{micrOMEGAs higg-sino}

\emph{Introduction.}--- Recent measurements have established that
nonbaryonic cold dark matter (CDM) makes up about one quarter of
the Universe's total energy budget \cite{Spergel:2006hy}. Although a
variety of candidates have been proposed \cite{Bertone:2004pz}, the
nature of this dark matter still remains a mystery.

One particularly attractive class of candidates is that of weakly
interacting massive particles (WIMPs), as gauge couplings and masses
of the electroweak symmetry breaking scale (i.e.\ presumably around
100 GeV) give the right order of magnitude for their thermal relic
abundance to explain the dark matter. Moreover, electroweak
precision tests indicate that the particle standard model Higgs
boson should be light: $m_h\!<\!144$ GeV at 95\% confidence level,
with a central value around 76 GeV \cite{lepewwg}. However, in the
standard model, the Higgs boson mass acquires quadratic ultraviolet
divergences, requiring fine-tuning to keep the Higgs mass low if no
new divergence-cancelling physics appears before the TeV energy
scale (the hierarchy problem). Therefore, new physics can be
expected to be found in upcoming high-energy experiments, such as
CERN's Large Hadron Collider (LHC).

The most studied scenario which provides both a symmetry to cancel
the quadratic divergences in the Higgs sector and a WIMP dark matter
candidate is supersymmetry. In the minimal supersymmetric standard
model, the lightest Higgs particle is constrained to be lighter than
$\sim$135 GeV \cite{PDG}, and although not excluded, some amount of
fine-tuning \cite{Barbieri:2006bg} is actually needed to fulfill the
experimental lower bound of 114 GeV \cite{lepewwg}. This has motivated
several studies on how to theoretically allow for an increase of the
Higgs mass both within supersymmetry and other extensions of the
standard model (see, e.g., \cite{Barbieri:2006bg,Barbieri:2006dq}
and references therein).

The \emph{Inert Doublet Model} (IDM) \cite{Barbieri:2006dq}
considered in this Letter is a very minimal extension of the
standard model -- an added second Higgs doublet $H_2$, with an
imposed unbroken discrete $Z_2$ symmetry that forbids its direct coupling to fermions (i.e.\ $H_2$ is \emph{inert}). 
In the IDM the standard model Higgs mass can be as high as about 500
GeV and still fulfill experimental precision tests
\cite{Barbieri:2006dq}.
Furthermore, conservation of the $Z_2$ parity implies that the
lightest inert Higgs particle ($H^0$) is stable and hence a good
dark matter candidate \cite{Ma:2006km,Barbieri:2006dq}.
Although the IDM does not solve the hierarchy problem, but
potentially only pushes the need for divergence cancelling physics
beyond the reach of upcoming accelerator searches such as the LHC
\cite{Barbieri:2006dq,Casas:2006bd}, it has the
advantage of providing a scalar WIMP 
dark matter candidate in a very
minimalistic way. 
An electrically neutral $H^0$ could therefore be used to represent a
larger class of scalar dark matter, similar to how the neutralino in
minimal supersymmetry often works as an archetype for supersymmetric
dark matter.
Actually, the origin of the IDM goes back to the 1970s
\cite{Deshpande:1977rw}
and has recently received new interest not only because of its
potential to allow for a high Higgs mass and a dark matter
candidate, but also for generating light neutrinos and leptogenesis
(see, e.g., \cite{Ma:2007yx} and references therein).
This gives strong reasons to study 
detection prospects for the inert Higgs.

A study of the relic density showed that $H^0$ can
constitute all dark matter if its mass is roughly $10-80$ GeV (or
above 500 GeV if parameters are tuned) \cite{LopezHonorez:2006gr}.
Direct detection studies show that the sensitivity of current
experiments is far too low, whereas future experiments could cover
more of the IDM parameter space \cite{
LopezHonorez:2006gr}.
The model can also produce
observable signals at the LHC \cite{Barbieri:2006dq}. As shown in \cite{LopezHonorez:2006gr},
indirect detection of the continuous gamma-ray spectrum
might be reachable with the upcoming 
GLAST experiment \cite{GLAST}.
However, this study was made for standard model Higgs masses of 120
and 200 GeV which, although giving higher gamma rates, deviates from
the motivation for the model of a raised Higgs mass
\cite{Barbieri:2006dq} and lacks distinctive detectable features.

An interesting aspect of the IDM is that the $H^0$ mass generically
has to be below the charged gauge boson mass, since the relatively
strong coupling to $W^+W^-$ would otherwise give a too low relic
density to explain the dark matter. Virtual $W$ bosons close to
threshold could, on the other hand, significantly enhance loop
processes producing monochromatic photons. We show here that this is
indeed correct
and study the dark matter ``smoking gun'' line signals from the final
states $\gamma\gamma$ and, when kinematically allowed, $Z\gamma$.
This, in combination with small tree-level
annihilation rates into fermions, makes the gamma lines the most
promising indirect detection signal.

Calculations of monochromatic gamma lines for both spin-1/2
\cite{Bergstrom:1997fh} and spin-1 \cite{Bergstrom:2004nr} dark
matter annihilation have been performed earlier.
The spin-0 dark matter candidate discussed here has
an even more promising gamma line rate.

\smallskip
\emph{The Inert Higgs Model and its Constraints.}--- The IDM
framework is an extension of the particle standard model with one
additional Higgs doublet $H_2$ and an unbroken $Z_2$ symmetry under
which $H_2$ is odd, $H_2\! \rightarrow \!-H_2$,
while all other fields are unchanged (even).
%
%
The
potential for such a model is
\cite{Barbieri:2006dq}
\begin{multline} \label{eq:potential}
V = \mu_1^2 \vert H_1\vert^2 + \mu_2^2 \vert H_2\vert^2  + \lambda_1 \vert H_1\vert^4 + \lambda_2 \vert H_2\vert^4\\
 + \lambda_3 \vert H_1\vert^2 \vert H_2 \vert^2
 + \lambda_4 \vert H_1^\dagger H_2\vert^2 + {\lambda_5} Re\!\left[(H_1^\dagger H_2)^2\right],
\end{multline}
where $H_1$ is the standard model Higgs doublet, and $\mu_i$ and
$\lambda_i$ are real parameters.

Besides the standard model Higgs particle ($h$), the physical states
derived from the inert doublet $H_2$ are two charged states
($H^\pm$) and two neutral: one CP-even ($H^0$) and one CP-odd
($A^0$), either of which is the dark matter candidate. For
definiteness we have chosen $H^0$ as the lightest inert particle,
although the role of $H^0$ and $A^0$ can in general be interchanged.
The masses are given by
\begin{eqnarray} \label{eq:masses}
 m_{h}^2   &=& -2 \mu_1^2\cr
 m_{H^0}^2 &=& \mu_2^2 + (\lambda_3 + \lambda_4 + \lambda_5) v^2\cr
 m_{A^0}^2 &=& \mu_2^2 + (\lambda_3 + \lambda_4 - \lambda_5) v^2\cr
 m_{H^\pm}^2 &=& \mu_2^2 +  \lambda_3 v^2\,,
\end{eqnarray}
where $v = m_{h}/\sqrt{4\lambda_1}$ is the vacuum expectation value
for $H_1$. 
As usual, the $W$ and $Z$ masses determine $v\sim 175$ GeV. No vacuum expectation value is allowed for $H_{2}$ because of the imposed unbroken $Z_{2}$ parity.

We apply the constraints in \cite{Barbieri:2006dq}
concerning vacuum stability, perturbativity, precision tests,
accelerator searches and naturalness \footnote{Because of arbitrariness
in defining \emph{naturalness}, we relax this constraint by a factor
of 2 compared to \protect\cite{Barbieri:2006dq}.} while using
direct detection and charged scalar mass bounds from
\cite{LopezHonorez:2006gr}. We also impose $m_{H^0}+m_{A^0}\gtrsim
m_Z$ to not be in conflict with data on the width of the $Z$
boson. Furthermore, in agreement with observations, 
we constrain the $H^0$ abundance to 0.094 $<\Omega_{CDM} h^2<$ 0.129
\cite{Spergel:2006hy}.
The $H^0$ relic density calculations
have been performed by
interfacing the standard numerical packages \textsc{FormCalc} \cite{FormCalc}
and \textsc{DarkSUSY} \cite{ds}.

\smallskip
\emph{Gamma-ray Lines from Inert Higgs Annihilations.}---
For $H^0$ masses below $m_W$, only annihilations into fermions
lighter than $m_{H^0}$ are accessible at tree level. The
annihilation rate is given by
 \begin{equation} \label{eq:ff}
   v_{\mathrm{rel}} \sigma_{f \bar f} = \frac{N_c \pi\alpha^2 m_f^2}{\sin^4{\theta_W} m_W^4}
   \frac{(1-\frac{4m_f^2}{s})^{3/2}(m_{H^0}^2-\mu_2^2)^2}{(s-m_{h}^2)^2+m_{h}^2\Gamma_h^2}\,,
 \end{equation}
where $N_c$ is a color factor (which equals 1 for leptons and 3 for
quarks), $\sqrt s$ is the center of mass energy, $\alpha$ the
fine-structure constant, $m_W$ the W boson mass, $\theta_W$ the weak
mixing angle, $\Gamma_h$ the decay width of $h$, and $m_f$ the final
state fermion mass.

The heaviest kinematically allowed fermion state will dominate the
tree-level annihilation channels.
The contributions to the continuum gamma-ray spectrum from
annihilating $H^0$ with masses below $m_W$ predominantly come from
the secondary gamma rays produced in the fragmentation of the bottom
quarks, mainly by the sequential production and decay of neutral
pions. Because of the harder gamma spectrum from the decay of
$\tau$-leptons they contribute significantly at the highest
energies, despite their much lower branching ratio. We use
\cite{PYTHIA} to calculate the shape of the
continuum spectra. 

We now turn to the main issue in this Letter -- the very important
line signals from direct annihilation of $H^0$ pairs into
$\gamma\gamma$ and $Z\gamma$. These spectral lines would show up as
characteristic dark matter fingerprints at the energies $m_{H^0}$
and $m_{H^0}\!-m^2_{Z}/4m_{H^0}$, respectively. The $Z\gamma$ line
might not be striclty monochromatic due to the Breit-Wigner width of the
Z mass, but can still be strongly peaked.
When the branching ratio
into $Z\gamma$ becomes significant the subsequent decay of the $Z$
boson significantly contributes to the continuum gamma-ray spectrum.
The full one-loop Feynman amplitudes were calculated by implementing
the IDM into the numerical \textsc{FormCalc} package
\cite{FormCalc}.

\begin{figure}[t]
 \includegraphics[width=0.99\columnwidth]{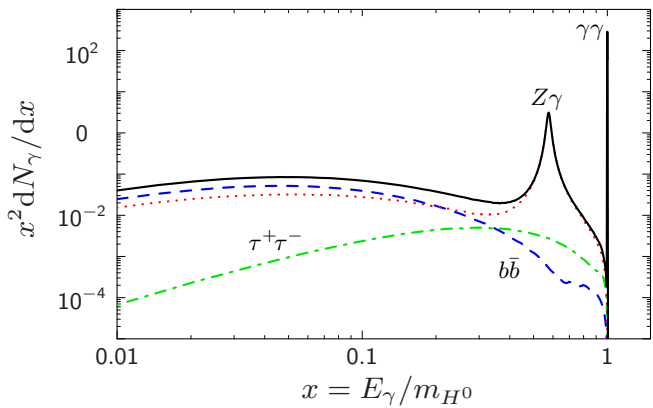}
\caption{\it The total differential photon distribution from
annihilations of an inert Higgs dark matter particle (solid line).
Shown separately are the contributions from $H^0 H^0 \rightarrow b
\bar b$ (dashed line), $\tau^+ \tau^-$ (dash-dotted line) and $Z \gamma$
(dotted line). This is for the benchmark model I in
Table~\ref{tab:benchmark}.}
 \label{fig:x2dNdx}
\end{figure}

Four IDM benchmark models are defined in Table~\ref{tab:benchmark},
where the two models III and IV are similar to those investigated
in \cite{LopezHonorez:2006gr}. Annihilation rates, branching ratios
and relic densities for these models are given in
Table~\ref{tab:benchmarkresults}. As an illustrative example,
Fig.~\ref{fig:x2dNdx} shows the predicted gamma spectrum for model
I.

\begin{figure}[t]
 \includegraphics[width=0.99\columnwidth]{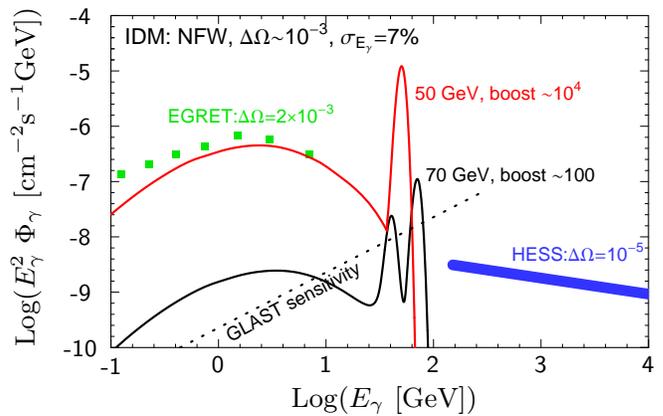}
\caption{\it Predicted gamma-ray spectra from the inert Higgs
benchmark models I and II 
as seen by GLAST (solid lines). The predicted gamma flux is from a
\mbox{$\Delta\Omega=10^{-3}$\! sr} region around the direction of
the galactic center assuming an NFW halo profile (with boost factors
as indicated in the figure) and convolved with a 7\! \% Gaussian
energy resolution. The boxes show EGRET data (which set an upper
limit for the continuum signal) and the thick line HESS data in the
same sky direction. The GLAST sensitivity (dotted line) is here
defined as 10 detected events within an effective exposure of
1~m$^2$yr within a relative energy range of $\pm$7\! \%.}
 \label{fig:GLAST}
\end{figure}

The spectral shape with its characteristic peaks in the hitherto
unexplored energy range between 30 and 100 GeV is ideal to search
for with the GLAST experiment \cite{Zaharijas:2006qb}. In
Fig.~\ref{fig:GLAST} this is illustrated by showing the predicted
fluxes from a $\Delta\Omega=10^{-3}$\! sr region around the
direction of the galactic center together with existing observations
in the same sky direction.
For simplicity, we assume a standard Navarro, Frenk and White (NFW)
density profile \cite{nfw} for the dark matter halo in our galaxy
($J \times \Delta\Omega \sim 1$ for $\Delta\Omega=10^{-3}$\! sr with
the notation of \cite{Cesarini:2003nr}). Processes such as adiabatic
compression could enhance the dark matter density significantly near
the galactic center (see, e.g., \cite{Bertone:2005hw}), and we
therefore allow our predicted flux to be scaled by a ``boost
factor''.
\begin{table}[b]
\vspace{-0.0cm} \caption{\it IDM benchmark models. (In units of
GeV.)}
 \begin{tabular*}{0.99\columnwidth}{@{\extracolsep{\fill}}ccccccc}
  \hline\hline
     Model      & $m_{h}$ & $m_{H^0}$ &   $m_{A^0}$&  $m_{H^\pm}$ &   $\mu_2$  &  $\lambda_2\times$1 GeV \\  \hline
      I  \      &  500        &  70   &     76     &   190      &   120      &      0.1                \\
      II \      &  500        &  50   &     58.5   &   170      &   120      &      0.1                \\
      III \     &  200        &  70   &     80     &   120      &   125      &      0.1                \\
      IV \      &  120        &  70   &     80     &   120      &    95      &      0.1                \\ \hline
 \end{tabular*}
 \label{tab:benchmark}
\end{table}
\begin{table}[b]
\vspace{-0.0cm}
 \caption{\it IDM benchmark model results.}
 \begin{tabular*}{0.99\columnwidth}{@{\extracolsep{\fill}}c@{\hspace{0.3cm}}c@{\hspace{0.5cm}}ccccc@{\hspace{0.5cm}}c}
  \hline\hline
 Model &$v\sigma^{v\rightarrow 0}_{tot}$&\multicolumn{5}{l}{Branching ratios [\%]:}                            & $\Omega_{\mathrm{CDM}} h^2$\\
       &[cm$^{3}$s$^{-1}$]              &$\gamma \gamma$&$Z \gamma$&$b \bar b$&$c \bar c$&$\tau^{+} \tau^{-}$ &                           \\\hline
   I   &  $1.6\times10^{-28}$           &      36     &   33   &   26   &  2   &   3            &   0.10                    \\
   II  &  $8.2\times10^{-29}$           &      29     &   0.6  &   60   &  4   &   7            &   0.10                    \\
   III &  $8.7\times10^{-27}$           &      2      &    2   &   81   &  5   &   9            &   0.12                    \\
   IV  &  $1.9\times10^{-26}$           &      0.04   &  0.1   &   85   &  5   &   10           &   0.11                    \\ \hline
 \end{tabular*}
 \label{tab:benchmarkresults}
\end{table}

The Energetic Gamma-Ray Experiment Telescope (EGRET) data, taken
from \cite{Cesarini:2003nr},
set an upper limit for the continuum part of our spectrum. For
example, for benchmark model II one finds that an optimistic, but
not necessarily unrealistic \cite{Bertone:2005hw}, boost of $10^{4}$
might be allowed.
In such cases, there would be a truly spectacular $\gamma\gamma$ line signal 
waiting for GLAST. However, to enable detection, boost factors of
such magnitudes are not necessary.
For $H^0$ masses closer to the $W$ threshold the $\gamma\gamma$
annihilation rates become even higher and in addition $Z\gamma$
production becomes important. In fact, these signals would
potentially be visible even without any boost at all (especially if
the background is low, as might be the case if the EGRET signal is
an galactic off-center source as indicated in \cite{Hooper:2002ru}).
Also shown in Fig.~\ref{fig:GLAST} is the data from the currently
operating air Cherenkov telescope HESS \cite{HESS}. One may notice
that future air Cherenkov telescopes with lower energy thresholds
will cover all of the interesting region for this dark matter
candidate.

\begin{figure}[!t]
 \includegraphics[width=0.99\columnwidth]{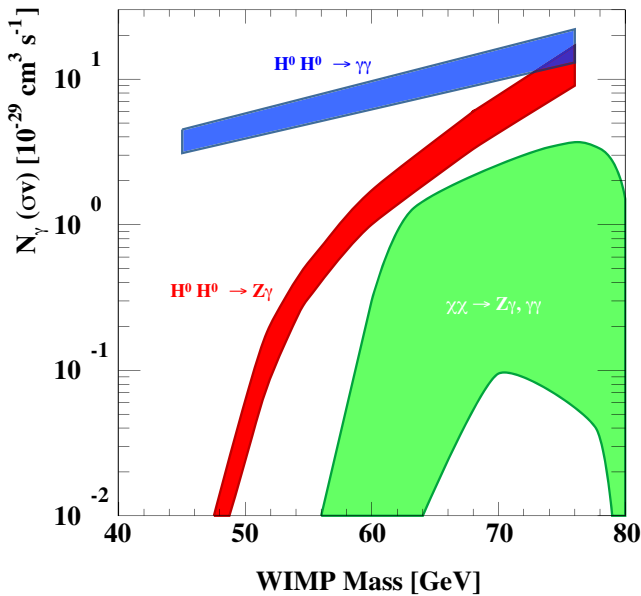}
\caption{\it Annihilation rates into gamma-ray lines
$2v\sigma_{\gamma\gamma}$ (upper band) and $v\sigma_{Z\gamma}$
(middle band) from the scan over the IDM parameter space. 
For comparison the lower-right region indicates the corresponding
results within the minimal supersymmetric standard model as obtained
with the \textsc{DarkSUSY} package \protect\cite{ds}.}
 \label{fig:scan}
\end{figure}

Finally, we have made a systematic parameter scan for $m_h=500$ GeV,
calculating the cross section into gamma lines. The previously
mentioned constraints allow us to scan the full parameter
space for dark matter masses below the $W$ threshold of 80 GeV. The
dependence on $m_{H^\pm}$ and $\lambda_2$ is small, and we set these
equal to $m_{H^0}+120$ GeV (to fulfill precision tests) and 0.1,
respectively. Importantly, one notes that the right relic density is
obtained with a significant amount of early Universe
coannihilations with the inert $A^0$ particle. The resulting
annihilation rates into $\gamma \gamma$ and $Z\gamma$ are shown in
Fig.~\ref{fig:scan}.
The lower and upper $m_{H^0}$ mass bounds come from the accelerator
constraints and the effect on the relic density by the opening of
the $W^+W^-$ annihilation channel, respectively.
For comparison, we show in the same figure the corresponding
annihilation rates for the neutralino ($\chi$) within the minimal
supersymmetric standard model.
The stronger line signal and smaller spread in the predicted IDM
flux are caused by the allowed unsuppressed coupling to $W$ pairs
that appear in contributing Feynman loop diagrams.
%

\smallskip
\emph{Summary and Conclusions.}--- In this Letter, we have
investigated the gamma-ray spectrum from the annihilation of the
inert Higgs dark matter
candidate $H^0$. 
In particular, we have focused on its striking gamma lines which
arise at the one-loop level and produce an exceptionally clear dark
matter signal.

The gamma line signals are particularly strong for this scalar dark
matter model mainly for two reasons: (1) The dark matter mass is just
below the kinematic threshold for $W$ production in the zero
velocity limit. (2) The dark matter candidate almost decouples from
fermions (i.e.,\ couples only via standard model Higgs exchange),
while still having ordinary gauge couplings to the gauge bosons.
In fact, these two properties could define a more general class of
models for which the IDM is an attractive archetype.
Despite small $H^0$ annihilation cross sections, coannihilations
with $A^0$ in the early Universe can bring the relic density of
$H^0$ into the correct range. The combination of the low (or even
completely vanishing) tree-level annihilation rates today and the
strong loop-level processes, due to unsuppressed couplings to
virtual $W$ bosons close to threshold, make the gamma lines the far
most dominant feature in the resulting gamma spectrum.

Absolute gamma-ray fluxes are, unfortunately, still hard to predict
due to the uncertainties in the structure of dark matter halos. It
might eventually be the spectral shape that enables a separation of
a dark matter signal and the background, in which case a gamma-line
would be a striking feature. We have shown that such signals in
the IDM are promising features to search for with the GLAST
satellite and with future air Cherenkov telescopes. One should bear
in mind that the best prospects for detection might not be in the
direction of the galactic center, but rather for other sources, such
as dwarf galaxies, smaller dark matter clumps or in the
extragalactic gamma-ray radiation, where the background is lower.

\bigskip
 M.G.~thanks Jan Conrad and Alexander Sellerholm for useful
discussions on the GLAST performance. M.G.~and E.L.~thank Thomas
Hahn for technical assistance with \textsc{FormCalc}. L.B.~and
J.E.~are grateful to the Swedish Science Research Council (VR) for
support.


\end{document}